\documentclass[preprint,10pt]{revtex4}
\usepackage{graphicx}
\def\beq{\begin{equation}}
\def\eeq{\end{equation}}
\def\bea{\begin{eqnarray}}
\def\eea{\end{eqnarray}}
\def\1{\c{c}}
\def\2{\c{C}}
\def\3{\u{g}}
\def\4{\u{G}}
\def\5{{\i}}
\def\6{\.{I}}
\def\7{\"{o}}
\def\8{\"{O}}
\def\9{\c{s}}
\def\0{\c{S}}
\def\*{\"{u}}
\def\,{\"{U}}
\begin{document}
\title{Scalar $f_{0}(980)$ meson effect in radiative $\phi\rightarrow
\pi^{+}\pi^{-}\gamma$ decay }
 \author{Ay\9e K\*\1\*karslan }
 \email{kucukarslan@metu.edu.tr}
\affiliation{Middle East Technical University, Physics Department. (06531), Ankara/Turkey}%
\author{Saime Solmaz}
 \email{skerman@balikesir.edu.tr}
\affiliation{Balikesir University, Physics Department.(10100),
Balikesir/Turkey}%
\date{\today}
\begin{abstract}
We study the effect of scalar-isoscalar $f_{0}(980)$ meson in the
mechanism of the radiative $\phi\rightarrow \pi^{+}\pi^{-}\gamma$
decay. A phenomenological approach is used to study this decay by
considering the contributions of $\sigma$-meson, $\rho$-meson and
$f_{0}(980)$-meson. The interference effects between different
contributions are analyzed and the branching ratio for this decay
is calculated. We observe that $f_0$ meson contribution is much
larger than the contributions of the other terms.
\end{abstract}
\maketitle
\section{Introduction}

Radiative decays of vector mesons offer the possibility of
investigating new physics features about the interesting mechanism
involved in these decays. One particular mechanism involves the
exchange of scalar mesons. The scalar mesons, isoscalar $\sigma$
and $f_{0}(980)$ and isovector $a_{0}(980)$, with vacuum quantum
numbers $J^{PC}=0^{++}$ are known to be crucial for a full
understanding of the low energy QCD phenomenology and the symmetry
breaking mechanisms in QCD. The scalar mesons have been a
persistent problem in hadron spectroscopy. In addition to the
identification of their nature, the role of scalar mesons in
hadronic processes is of extreme importance and the study of
radiative decays of vector mesons may provide insights about their
role.

In particular, radiative $\phi$ meson decays,
$\phi\rightarrow\pi\pi\gamma$ and
$\phi\rightarrow\pi^{0}\eta\gamma$, can play a crucial role in the
clarification of the structure and properties of scalar
$f_{0}(980)$ and $a_{0}(980)$ mesons since these decays primarily
proceed through processes involving scalar resonances such as
$\phi\rightarrow f_{0}(980)\gamma$ and $\phi\rightarrow
a_{0}(980)\gamma$, with the subsequent decays into $\pi\pi\gamma$
and $\pi^{0}\eta\gamma$ \cite{R1,R2}. Achasov and Ivanchenko
\cite{R1} showed that if the $f_{0}(980)$ and $a_{0}(980)$
resonances are four-quark $(q^{2}\bar{q}^{2})$ states the
processes $\phi\rightarrow f_{0}(980)\gamma$ and $\phi\rightarrow
a_{0}(980)\gamma$ are dominant and enhance the decays
$\phi\rightarrow\pi\pi\gamma$ and
$\phi\rightarrow\pi^{0}\eta\gamma$ by at least an order of
magnitude over the results predicted by the Wess-Zumino terms.
Then Close et al. \cite{R2} noted that the study of the scalar
states in $\phi\rightarrow S\gamma$, where $S=f_{0}~ or~ a_{0}$,
may offer unique insights into the nature of the scalar mesons.
They have shown that although the transition rates
$\Gamma(\phi\rightarrow f_{0}\gamma)$ and $\Gamma(\phi\rightarrow
a_{0}\gamma)$ depend on the unknown dynamics, the ratio of the
decay rates $\Gamma(\phi\rightarrow
a_{0}\gamma)/\Gamma(\phi\rightarrow f_{0}\gamma)$ provides an
experimental test which distinguishes between alternative
explanations of their structure. On the experimental side, the
Novosibirsk CMD-2  \cite{R3,R4} and SND  \cite{R5} collaborations
give the following branching ratios for
$\phi\rightarrow\pi^{+}\pi^{-}\gamma$ and
$\phi\rightarrow\pi^{0}\eta\gamma$ decays:
$BR(\phi\rightarrow\pi^{+}\pi^{-}\gamma)=(0.41\pm0.12\pm0.04)\times10^{-4}$
\cite{R3},
$BR(\phi\rightarrow\pi^{0}\eta\gamma)=(0.90\pm0.24\pm0.10)\times10^{-4}$
\cite{R4},
$BR(\phi\rightarrow\pi^{0}\eta\gamma)=(0.88\pm0.14\pm0.09)\times10^{-4}$
\cite{R5}, where the first error is statistical and the second one
is systematic.

Theoretically, the role of $f_{0}(980)$-meson in the radiative
decay processes $\phi\rightarrow\pi\pi\gamma$ was also
investigated by Achasov et al. \cite{R6}. They calculated the
branching ratio for this decay by considering only
$f_{0}(980)$-meson contribution. They used two different models of
$f_{0}(980)$-meson: the four-quark model and $K\bar{K}$ molecular
model. In the four-quark model they obtained the value for the
branching ratio as $BR(\phi\rightarrow
f_{0}\gamma\rightarrow\pi\pi\gamma)=2.3\times10^{-4}$ and in case
of the $K\bar{K}$ molecular model, the branching ratio was
$BR(\phi\rightarrow
f_{0}\gamma\rightarrow\pi\pi\gamma)=1.7\times10^{-5}$. Later,
Marco et al. considered the radiative $\phi$ meson decays
\cite{R7} as well as other radiative vector meson decays within
the framework of chiral unitary theory developed earlier by Oller
\cite{R8}. They obtained the result
$BR(\phi\rightarrow\pi^{+}\pi^{-}\gamma)=1.6\times10^{-4}$ for the
branching ratio of the $\phi\rightarrow\pi^{+}\pi^{-}\gamma$ decay
and emphasized that the branching ratio for
$\phi\rightarrow\pi^{+}\pi^{-}\gamma$ decay is twice the one for
$\phi\rightarrow\pi^{0}\pi^{0}\gamma$ decay. Recently, the
radiative $\phi\rightarrow\pi^{0}\pi^{0}\gamma$ decay, where the
scalar $f_{0}(980)$-meson plays an important role was studied by
G\7kalp and Y\5lmaz \cite{R9} within the framework of a
phenomenological approach in which the contributions of
$\sigma$-meson, $\rho$-meson and $f_{0}$-meson are considered.
They analyzed the interference effects between different
contributions. Their analysis showed that $f_{0}(980)$-meson
amplitude makes a substantial contribution to the branching ratio
of this decay. Furthermore, recently Escribano has been studied
the scalar meson exchange in $V\rightarrow\pi^{0}\pi^{0}\gamma$
decays \cite{R10}. He discussed the scalar contributions to the
$\phi\rightarrow\pi^{0}\pi^{0}\gamma$,
$\phi\rightarrow\pi^{0}\eta\gamma$ and
$\rho^{0}\rightarrow\pi^{0}\pi^{0}\gamma$ decays in the framework
of the linear sigma model ($L \sigma M$). He obtained the result
$BR(\phi\rightarrow\pi^{0}\pi^{0}\gamma)=1.16\times10^{-4}$ for
the branching ratio of the $\phi\rightarrow\pi^{0}\pi^{0}\gamma$
decay and noted that, the branching ratio for this decay is
dominated by $f_{0}(980)$ meson amplitude.

In this work, we study the radiative vector meson decay
$\phi\rightarrow\pi^{+}\pi^{-}\gamma$ to investigate the role of
the scalar $f_{0}(980)$ meson and to extract the relevant
information on the properties of this scalar meson. Theoretically,
the radiative $\phi\rightarrow\pi^{+}\pi^{-}\gamma$ decay has not
been studied extensively up to now. One of the rare studies of
this decay was by Marco et al. \cite{R7} who neglected the
contributions coming from intermediate vector meson states.
Therefore, this decay should be reconsidered and the VMD amplitude
should be added to the $f_{0}$-meson and $\sigma$-meson
amplitudes.

\section{Formalism}

We study the radiative decay $\phi\rightarrow
\pi^{+}\pi^{-}\gamma$ within the framework of a phenomenological
approach in which the contributions of $\sigma$-meson,
$\rho$-meson and $f_{0}$-meson are considered and we do not make
any assumption about the structure of the $f_0$ meson. In our
phenomenological approach we describe the $\phi KK$-vertex by the
effective Lagrangian
\begin{eqnarray}\label{e1}
 {\cal L}^{eff.}_{\phi KK}=-ig_{\phi KK}\phi^{\mu}
 \left(K^{+}\partial_{\mu}K^{-}-K^{-}\partial_{\mu}K^{+}\right)~~,
\end{eqnarray}
and for the $f_{0} KK$-vertex we use the phenomenological
Lagrangian
\begin{eqnarray}\label{e2}
 {\cal L}^{eff.}_{f_{0} KK}=g_{f_{0} KK}M_{f_{0}}K^{+}K^{-}f_{0}~~.
\end{eqnarray}
\begin{figure}[t]
$\left. \right.$ \vskip -2cm  \includegraphics{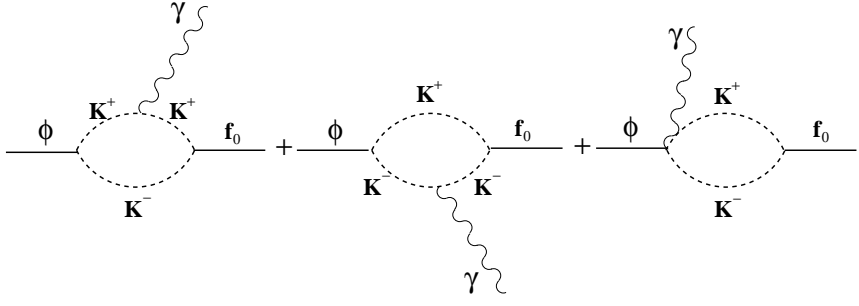} \vskip 7cm
\caption{Feynman diagrams for the decay $\phi\rightarrow
f_{0}\gamma$}
\end{figure}
The effective Lagrangians for the $\phi KK$- and $f_{0}
KK$-vertices also serve to define the coupling constants $g_{\phi
KK}$ and $g_{f_{0}KK}$ respectively. The decay width for the
$\phi\rightarrow K^{+}K^{-}$ decay is obtained from the Lagrangian
given in Eq. 1 and this decay width is
\begin{eqnarray}\label{e3}
 \Gamma(\phi\rightarrow K^{+}K^{-})=\frac{g^{2}_{\phi KK}}{48 \pi}
  M_{\phi} \left[1-\left(\frac{2M_{K}}{M_{\phi}}\right)^{2}\right]^{3/2}~~.
\end{eqnarray}
We then obtain the coupling constant $g_{\phi KK}$ from the
experimental partial width \cite{R11} of the radiative decay
$\phi\rightarrow K^{+}K^{-}$ as $g_{\phi KK}=(4.59\pm 0.05)$. The
amplitude of the radiative decay $\phi\rightarrow f_{0}\gamma$ is
obtained from the diagrams shown in Fig. 1 where the last diagram
assures gauge invariance \cite{R1,R12}. This amplitude is
\begin{eqnarray}\label{e4}
 {\cal M }\left(\phi\rightarrow f_{0}\gamma\right)=-\frac{1}{2\pi^{2}M_{K}^{2}}\left(g_{f_{0}KK}M_{f_{0}}\right)
 \left(eg_{\phi KK}\right)I(a,b)\left[\epsilon\cdot u ~k\cdot p-\epsilon\cdot p~ k\cdot u\right]~
\end{eqnarray}
where $(u,p)$ and $(\epsilon,k)$ are the polarizations and
four-momenta of the $\phi$ meson and the photon respectively, and
also $a=M_{\phi}^{2}/M_{K}^{2}$, $b=M_{f_{0}}^{2}/M_{K}^{2}$. The
$I(a,b)$ function has been calculated in different contexts
\cite{R2,R8,R13} and is defined as
\begin{eqnarray}\label{e5}
 I(a,b)=\frac{1}{2(a-b)} -\frac{2}{(a-b)^{2}}
 \left [f(\frac{1}{b})-f(\frac{1}{a})\right ] +\frac{a}{(a-b)^{2}}\left [
 g(\frac{1}{b})-g(\frac{1}{a})\right ]~~,
\end{eqnarray}
where
\begin{eqnarray}\label{e6}
&&f(x)=\left \{ \begin{array}{rr}
              -\left [ \arcsin (\frac{1}{2\sqrt{x}})\right ]^{2}~,& ~~x>\frac{1}{4} \\
               \frac{1}{4}\left [ \ln (\frac{\eta_{+}}{\eta_{-}})-i\pi\right]^{2}~, & ~~x<\frac{1}{4}
             \end{array} \right.\nonumber \\
&& \nonumber \\
&&g(x)=\left \{ \begin{array}{rr}
              (4x-1)^{\frac{1}{2}} \arcsin(\frac{1}{2\sqrt{x}})~, & ~~ x>\frac{1}{4} \\
              \frac{1}{2}(1-4x)^{\frac{1}{2}}\left [\ln (\frac{\eta_{+}}{\eta_{-}})-i\pi \right ]~, & ~~ x<\frac{1}{4}
            \end{array} \right.\nonumber \\
&& \nonumber \\
&&\eta_{\pm}=\frac{1}{2x}\left [ 1\pm(1-4x)^{\frac{1}{2}}\right ]
~.
\end{eqnarray}
Then, the decay rate for the $\phi\rightarrow f_{0}\gamma$ decay
is
\begin{eqnarray}\label{e7}
 \Gamma(\phi\rightarrow f_{0}\gamma)=\frac{\alpha}{6 (2\pi)^4}
 \frac{M_{\phi}^{2}-M_{f_{0}}^{2}}{M_{\phi}^{3}}g^{2}_{\phi KK}
 \left(g_{f_{0}KK} M_{f_{0}}\right)^2
 \left|(a-b)I(a,b)\right|^{2}~~.
\end{eqnarray}
Utilizing the experimental value for the branching ratio
$BR(\phi\rightarrow f_{0}\gamma)=(3.4\pm 0.4)\times 10^{-4}$ for
the decay $\phi\rightarrow f_{0}\gamma$ \cite{R11}, we determine
the coupling constant $g_{f_{0}KK}$ as $g_{f_{0}KK}=(4.13\pm
1.42)$. In our calculation, we assume that the radiative decay
$\phi\rightarrow \pi^{+}\pi^{-}\gamma$ proceeds through the
reactions $\phi\rightarrow\sigma\gamma\rightarrow
\pi^{+}\pi^{-}\gamma$,
$\phi\rightarrow\rho^{\mp}\pi^{\pm}\rightarrow
\pi^{+}\pi^{-}\gamma$ and $\phi\rightarrow f_{0}\gamma\rightarrow
\pi^{+}\pi^{-}\gamma$. Therefore, our calculation is based on the
Feynman diagrams shown in Fig. 2. For the
$\phi\sigma\gamma$-vertex, we use the effective Lagrangian
\begin{figure}[t]
$\left. \right.$ \vskip 1.5cm  \includegraphics{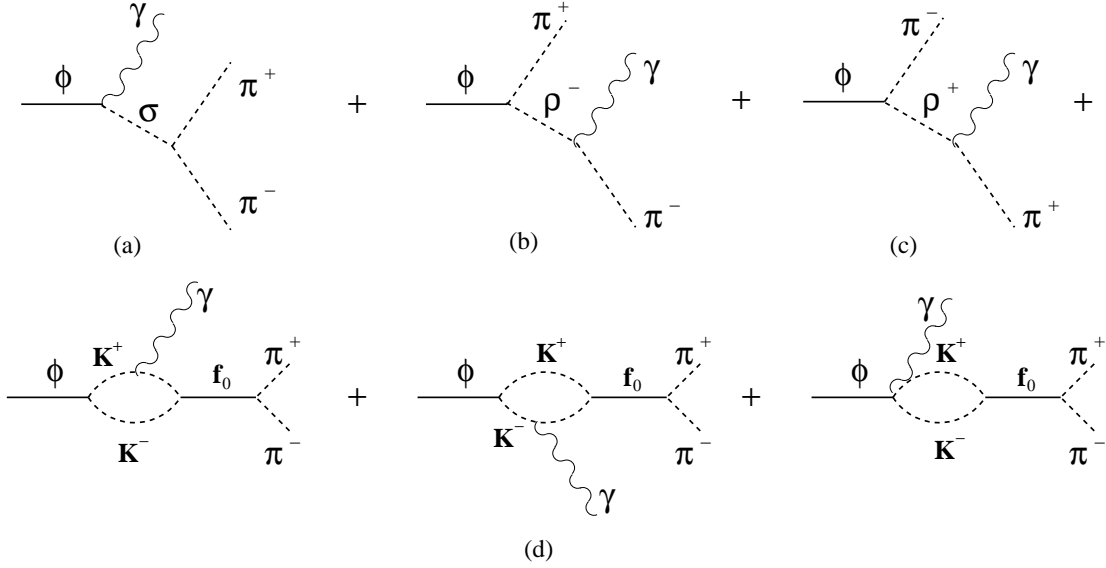} \vskip 7cm
\caption{Feynman diagrams for the decay $\phi\rightarrow
\pi^+\pi^-\gamma$}
\end{figure}
\begin{eqnarray}\label{e8}
 {\cal L}^{eff.}_{\phi\sigma\gamma}=\frac{e}{M_{\phi}}g_{\phi\sigma\gamma}
 [\partial^{\alpha}\phi^{\beta}\partial_{\alpha}A_{\beta}
 -\partial^{\alpha}\phi^{\beta}\partial_{\beta}A_{\alpha}]\sigma~~,
\end{eqnarray}
which also defines the coupling constant $g_{\phi\sigma\gamma}$.
The coupling constant $g_{\phi\sigma\gamma}$ is determined by
G\7kalp and Y\5lmaz \cite{R9} as $g_{\phi\sigma\gamma}=(0.025\pm
0.009)$ using the experimental value of the branching ratio for
the radiative decay $\phi\rightarrow \pi^{0}\pi^{0}\gamma$
\cite{R14}. For the $\sigma\pi\pi$-vertex we use the effective
Lagrangian
\begin{eqnarray}\label{e9}
 {\cal L}^{eff.}_{\sigma\pi\pi}=\frac{1}{2}g_{\sigma\pi\pi}
 M_{\sigma}\vec{\pi}\cdot\vec{\pi}\sigma~~.
\end{eqnarray}
The decay width of the $\sigma$-meson that results from this
effective Lagrangian is given as
\begin{eqnarray}\label{e10}
 \Gamma_{\sigma}\equiv\Gamma(\sigma\rightarrow\pi\pi)=
 \frac{g^{2}_{\sigma\pi\pi}}{4\pi}\frac{3M_{\sigma}}{8}
 \left[1-\left(\frac{2M_{\pi}}{M_{\sigma}}\right)^{2}\right]^{1/2}~~.
\end{eqnarray}
For given values of $M_{\sigma}$ and $\Gamma_{\sigma}$, we use
this expression to determine the coupling constant
$g_{\sigma\pi\pi}$. Therefore, using the experimental values for
$M_{\sigma}$ and $\Gamma_{\sigma}$ \cite{R15}, given as
$M_{\sigma}=(478\pm 17)~MeV$ and $\Gamma_{\sigma}=(324\pm
21)~MeV$, we obtain the coupling constant
$g_{\sigma\pi\pi}=(5.25\pm 0.32)$. The $\phi\rho\pi$-vertex is
conventionally described by the effective Lagrangian
\begin{eqnarray}\label{e11}
 {\cal L}^{eff.}_{\phi\rho\pi}=\frac{g_{\phi\rho\pi}}{M_\phi}
 \epsilon^{\mu\nu\alpha\beta}\partial_{\mu}\phi_{\nu}
 \partial_{\alpha}\vec{\rho_{\beta}}\cdot\vec{\pi}~~.
\end{eqnarray}
The coupling constant $g_{\phi\rho\pi}$ is calculated as
$g_{\phi\rho\pi}=(0.811\pm 0.081)~GeV^{-1}$ by Achasov and Gubin
\cite{R16} using the data on the decay
$\phi\rightarrow\rho\pi\rightarrow\pi^{+}\pi^{-}\pi^{0}$
\cite{R11}. For the $\rho\pi\gamma$-vertex the effective
Lagrangian
\begin{eqnarray}\label{e12}
 {\cal L}^{eff.}_{\rho\pi\gamma}=\frac{e}{M_{\rho}}g_{\rho\pi\gamma}
 \epsilon^{\mu\nu\alpha\beta}\partial_{\mu}\vec{\rho_{\nu}}\cdot\vec{\pi}~\partial_{\alpha}A_{\beta}~~,
\end{eqnarray}
is used. At present there is a discrepancy between the
experimental widths of the $\rho^0\rightarrow\pi^{0}\gamma$ and
$\rho^+\rightarrow\pi^{+}\gamma$ decays. We use the experimental
rate for the decay $\rho^0\rightarrow\pi^{0}\gamma$ \cite{R11} to
extract the coupling constant $g_{\rho\pi\gamma}$ as
$g_{\rho\pi\gamma}=(0.69\pm 0.35)$ since the experimental value
for the decay rate of $\phi\rightarrow\pi^{0}\pi^{0}\gamma$ was
used by G\7kalp and Y\5lmaz \cite{R9} to estimate the coupling
constant $g_{\phi\sigma\gamma}$. Finally, the $f_{0}\pi\pi$-vertex
is described conventionally by the effective Lagrangian
\begin{eqnarray}\label{e13}
 {\cal L}^{eff.}_{f_{0}\pi\pi}=\frac{1}{2}g_{f_{0}\pi\pi}
 M_{f_{0}}\vec{\pi}\cdot\vec{\pi}f_{0}~~.
 \end{eqnarray}
In our calculation of the invariant amplitude, we make the
replacement $q^{2}-M^{2}\rightarrow q^{2}-M^{2}+iM\Gamma $, where
$q$ and $M$ are four-momentum and mass of the virtual particles
respectively, in $\rho$-, $\sigma$- and $f_{0}$- propagators in
order to take into account the finite widths of these unstable
particles and use the experimental value $\Gamma_{\rho}=(150.2\pm
0.8)~MeV$ \cite{R11} for $\rho$-meson. However, since the mass
$M_{f_{0}}=980~MeV$ of $f_{0}$-meson is very close to the
$K^{+}K^{-}$ threshold this gives rise to a strong energy
dependence on the width of the $f_{0}$-meson and to include this
energy dependence different expressions for $\Gamma_{f_{0}}$ can
be used. First option is to use an energy dependent width for
$f_{0}$
\begin{eqnarray}\label{e14}
 \Gamma_{f_{0}}(q^{2})=\Gamma_{\pi\pi}^{f_{0}}(q^{2})~\theta\left(\sqrt{q^{2}}-2M_{\pi}\right)
 +\Gamma_{K\overline{K}}^{f_{0}}(q^{2})~\theta\left(\sqrt{q^{2}}-2M_{K}\right)~~,
\end{eqnarray}
where $q^{2}$ is the four-momentum square of the virtual
$f_{0}$-meson and the width $\Gamma_{\pi\pi}^{f_{0}}(q^{2})$ is
given as
\begin{eqnarray}\label{e15}
 \Gamma_{\pi\pi}^{f_{0}}(q^{2})=\Gamma_{\pi\pi}^{f_{0}}~\frac{M_{f_{0}}^2}{q^2}
 \sqrt{\frac{q^{2}-4M_{\pi}^{2}}{M_{f_{0}}^{2}-4M_{\pi}^{2}}}~~.
\end{eqnarray}
We use the experimental value for $\Gamma_{\pi\pi}^{f_{0}}$ as
$\Gamma_{\pi\pi}^{f_{0}}=40-100~MeV$ \cite{R11}. The width
$\Gamma_{K\overline{K}}^{f_{0}}(q^{2})$ is given by a similar
expression as for $\Gamma_{\pi\pi}^{f_{0}}(q^{2})$. Another and
widely accepted option is the work of Flatt\'{e} \cite{R17}. In
his work, the expression for
$\Gamma_{K\overline{K}}^{f_{0}}(q^{2})$ is extended below the
$K\overline{K}$ threshold where $\sqrt{q^{2}-4M_{K}^{2}}$ is
replaced by $i\sqrt{4M_{K}^{2}-q^{2}}$ so
$\Gamma_{K\overline{K}}^{f_{0}}(q^{2})$ becomes purely imaginary.
However in our work, we take into account both options. The
invariant amplitude ${\cal M}(E_{\gamma}, E_{1})$ is expressed as
${\cal M}(E_{\gamma}, E_{1})={\cal M}_{a}+{\cal M}_{b}+{\cal
M}_{c}+{\cal M}_{d}$ where ${\cal M}_{a}$, ${\cal M}_{b}$, ${\cal
M}_{c}$ and ${\cal M}_{d}$ are the invariant amplitudes resulting
from the diagrams $(a)$, $(b)$, $(c)$ and $(d)$ in Fig. 2
respectively. Therefore, the interference between different
reactions contributing to the decay $\phi\rightarrow
\pi^{+}\pi^{-}\gamma$ is taken into account. The differential
decay probability for an unpolarized $\phi$-meson at rest is given
as
\begin{eqnarray}\label{e16}
\frac{d\Gamma}{dE_{\gamma}dE_{1}}=\frac{1}{(2\pi)^{3}}~\frac{1}{8M_{\phi}}~
\mid {\cal M}\mid^{2}~~,
\end{eqnarray}
where E$_{\gamma}$ and E$_{1}$ are the photon and pion energies
respectively. We perform an average over the spin states of
$\phi$-meson and a sum over the polarization states of the photon.
The decay width $\Gamma(\phi\rightarrow\pi^{+}\pi^{-}\gamma)$ is
then obtained by integration
\begin{eqnarray}\label{e17}
\Gamma=\int_{E_{\gamma,min.}}^{E_{\gamma,max.}}dE_{\gamma}
       \int_{E_{1,min.}}^{E_{1,max.}}dE_{1}\frac{d\Gamma}{dE_{\gamma}dE_{1}}~~,
\end{eqnarray}
where the minimum photon energy is E$_{\gamma, min.}=0$ and the
maximum photon energy is given as
$E_{\gamma,max.}=(M_{\phi}^{2}-4M_{\pi}^{2})/2M_{\phi}=471.8~MeV$.
The maximum and minimum values for the pion energy E$_{1}$ are
given by
\begin{eqnarray}\label{e18}
\frac{1}{2(2E_{\gamma}M_{\phi}-M_{\phi}^{2})} [
-2E_{\gamma}^{2}M_{\phi}+3E_{\gamma}M_{\phi}^{2}-M_{\phi}^{3}
 ~~~~~~~~~~~~~~~~~~~~~~~~~~~~ \nonumber \\
\pm  E_{\gamma}\sqrt{(-2E_{\gamma}M_{\phi}+M_{\phi}^{2})
       (-2E_{\gamma}M_{\phi}+M_{\phi}^{2}-4M_{\pi}^{2})}~]~~.
\end{eqnarray}


\section{Results and Discussion}

In order to determine the coupling constant $g_{f_{0}\pi\pi}$, we
choose for the $f_{0}$-meson parameters the values
$M_{f_{0}}=980~MeV$ and $\Gamma_{f_{0}}=(70\pm30)~MeV$. Therefore,
through the decay rate that results from the effective Lagrangian
given in Eq. 13 we obtain the coupling constant $g_{f_{0}\pi\pi}$
as $g_{f_{0}\pi\pi}=(1.58\pm0.30)$. If we use the form for
$\Gamma^{f_{0}}_{K\bar{K}}(q^2)$, proposed by Flatt\'{e}
\cite{R17}, the desired enhancement in the invariant mass spectrum
appears in its central part rather than around the $f_{0}$ pole.
Bramon et al. \cite{R18} also encountered a similar problem in
their study of the effects of  the $a_{0}(980)$ meson in the
$\phi\rightarrow\pi^0 \eta\gamma$ decay. Therefore, in the
analysis which we present below for $\Gamma_{f_{0}}(q^2)$ we use
the form given in Eq. 14. The invariant mass distribution
$dB/dM_{\pi\pi}=(M_{\pi\pi}/M_{\phi})dB/dE_{\gamma}$ for the
radiative decay $\phi\rightarrow \pi^+\pi^-\gamma$ is plotted in
Fig. 3 as a function of the invariant mass $M_{\pi\pi}$ of
$\pi^+\pi^-$ system. In this figure we indicate the contributions
coming from different reactions
$\phi\rightarrow\sigma\gamma\rightarrow\pi^+\pi^-\gamma$,
$\phi\rightarrow\rho^{\mp}\pi^{\pm}\rightarrow\pi^+\pi^-\gamma$
and $\phi\rightarrow f_{0}\gamma\rightarrow\pi^+\pi^-\gamma$ as
well as the contribution of the total amplitude which includes the
interference terms as well. It is clearly seen from Fig. 3 that
the spectrum for the decay $\phi\rightarrow \pi^+\pi^-\gamma$ is
dominated by the $f_{0}$-amplitude. On the other hand the
contribution coming from $\sigma$-amplitude can only be noticed in
the region $M_{\pi\pi}<0.7~GeV$ through interference effects.
Likewise $\rho$-meson contribution can be seen in the region
$M_{\pi\pi}<0.8~GeV$ so we can say that the contribution of the
$f_{0}$-term is much larger than the contributions of the
$\sigma$-term and $\rho$-term as well as the contribution of the
total interference term having opposite sign.
\begin{figure}[t]
$\left. \right.$ \vskip 8.4cm  \includegraphics{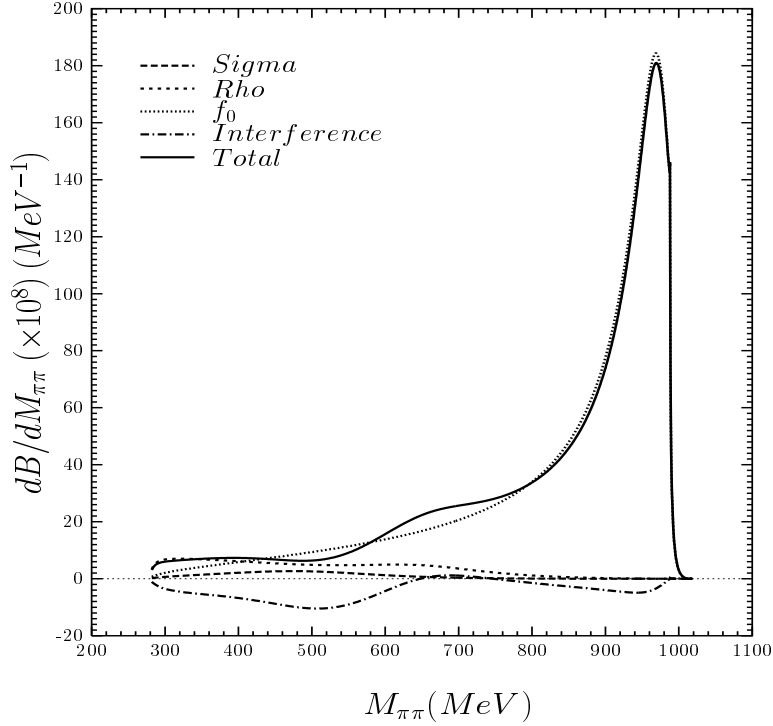} \vskip 2cm
\caption{ The $\pi\pi$ invariant mass spectrum for the decay
$\phi\rightarrow \pi^+\pi^-\gamma$. The contributions of different
terms are indicated.}
\end{figure}
The dominant $f_{0}$-term characterizes the invariant mass
distribution  in the region where $M_{\pi\pi}>0.7~GeV$. In our
study contributions of different amplitudes to the branching ratio
of the decay $\phi\rightarrow \pi^+\pi^-\gamma$ are
$BR(\phi\rightarrow
f_{0}\gamma\rightarrow\pi^+\pi^-\gamma)=2.54\times 10^{-4}$,
$BR(\phi\rightarrow\sigma\gamma\rightarrow\pi^+\pi^-\gamma)=0.07\times
10^{-4}$,
$BR(\phi\rightarrow\rho^{\mp}\pi^{\pm}\rightarrow\pi^+\pi^-\gamma)=0.26\times
10^{-4}$,
$BR(\phi\rightarrow(f_{0}\gamma+\pi^{\pm}\rho^{\mp})\rightarrow\pi^+\pi^-\gamma)=2.74\times
10^{-4}$,
$BR(\phi\rightarrow(f_{0}\gamma+\sigma\gamma)\rightarrow\pi^+\pi^-\gamma)=2.29\times
10^{-4}$ and for the total interference term
$BR(\textrm{interference})=-0.29\times 10^{-4}$. We then calculate
the total branching ratio as $BR(\phi\rightarrow
\pi^+\pi^-\gamma)=2.57\times 10^{-4}$. Our result is twice the
theoretical result for $\phi\rightarrow \pi^0\pi^0\gamma$ decay,
obtained by G\7kalp and Y\5lmaz \cite{R9}, as it should be. They
obtained the following values: $BR(\phi\rightarrow
f_{0}\gamma\rightarrow\pi^0\pi^0\gamma)=1.29\times 10^{-4}$,
$BR(\phi\rightarrow\sigma\gamma\rightarrow\pi^0\pi^0\gamma)=0.04\times
10^{-4}$,
$BR(\phi\rightarrow\rho^{0}\pi^{0}\rightarrow\pi^0\pi^0\gamma)=0.14\times
10^{-4}$,
$BR(\phi\rightarrow(f_{0}\gamma+\pi^{0}\rho^{0})\rightarrow\pi^0\pi^0\gamma)=1.34\times
10^{-4}$,
$BR(\phi\rightarrow(f_{0}\gamma+\sigma\gamma)\rightarrow\pi^0\pi^0\gamma)=1.16\times
10^{-4}$ and $BR\textrm{(interference)}=-0.25\times 10^{-4}$.
Moreover, our calculation for the branching ratio of the radiative
decay $\phi\rightarrow\pi^+\pi^-\gamma$ is nearly twice the value
for the branching ratio of the radiative decay
$\phi\rightarrow\pi^0\pi^0\gamma$ that was obtained by Achasov and
Gubin \cite{R16}. Besides, $\phi\rightarrow\pi^{+}\pi^{-}\gamma$
decay was considered by Marco et al. \cite{R7} in the framework of
unitarized chiral perturbation theory. The branching ratio for
$\phi\rightarrow\pi^{+}\pi^{-}\gamma$, they obtained, was
$BR(\phi\rightarrow\pi^{+}\pi^{-}\gamma)=1.6\times 10^{-4}$ and
for $\phi\rightarrow\pi^{0}\pi^{0}\gamma$ was
$BR(\phi\rightarrow\pi^{0}\pi^{0}\gamma)=0.8\times 10^{-4}$. As we
mentioned above, they noted that the branching ratio for
$\phi\rightarrow\pi^{0}\pi^{0}\gamma$ is one half of
$\phi\rightarrow\pi^{+}\pi^{-}\gamma$. Therefore our calculation
for the branching ratio of $\phi\rightarrow\pi^{+}\pi^{-}\gamma$
decay is in accordance with the theoretical expectations. A
similar relation can be seen between the decay rates of
$\omega\rightarrow\pi^{+}\pi^{-}\gamma$ and
$\omega\rightarrow\pi^{0}\pi^{0}\gamma$ \cite{R19}. It was noticed
that
$\Gamma(\omega\rightarrow\pi^{0}\pi^{0}\gamma)=1/2\Gamma(\omega\rightarrow\pi^{+}\pi^{-}\gamma)$
and the factor $1/2$ is a result of charge conjugation invariance
to order $\alpha$ which imposes pion pairs of even angular
momentum. The experimental value of the branching ratio for
$\phi\rightarrow\pi^{+}\pi^{-}\gamma$, measured by Akhmetshin et
al., is
$BR(\phi\rightarrow\pi^{+}\pi^{-}\gamma)=(0.41\pm0.12\pm0.04)\times
10^{-4}$ \cite{R3}. So the value of the branching ratio that we
obtained is approximately six times larger than the value of the
measured branching ratio. As it was stated by Marco et al.
\cite{R7}, we should not compare our calculation for the branching
ratio of the radiative decay $\phi\rightarrow\pi^{+}\pi^{-}\gamma$
directly with experiment since the experiment is done using the
reaction $e^{+}e^{-}\rightarrow\phi\rightarrow\pi^+\pi^-\gamma$,
which interferes with the off-shell $\rho$ dominated amplitude
coming from the reaction
$e^{+}e^{-}\rightarrow\rho\rightarrow\pi^+\pi^-\gamma$ \cite{R20}.
Also the result in \cite{R3} is based on model dependent
assumptions.
\section{Acknowledgement}
We thank A. G\7kalp and O. Y\5lmaz for their invaluable comments
and suggestions during this work.


\begin{thebibliography}{99}

\bibitem{R1}   N. N. Achasov, V. N. Ivanchenko, Nucl. Phys.  {\bf B315}, 465 (1989).
\bibitem{R2}   F. E. Close, N. Isgur, S. Kumona, Nucl. Phys.  {\bf B389}, 513 (1993).
\bibitem{R3}   R. R. Akhmetshin et al., Phys. Lett.  {\bf B462}, 371 (1999).
\bibitem{R4}   R. R. Akhmetshin et al., Phys. Lett.  {\bf B462}, 380 (1999).
\bibitem{R5}   M. N. Achasov et al., Phys. Lett.  {\bf B479}, 53 (2000).
\bibitem{R6}   N. N. Achasov, V. V. Gubin and E. P. Solodov, Phys. Rev. {\bf D55}, 2672 (1997).
\bibitem{R7}   E. Marco, S. Hirenzaki, E. Oset and H. Toki, Phys. Lett. {\bf B470}, 20 (1999).
\bibitem{R8}   J. A. Oller, Phys. Lett. {\bf B426}, 7 (1998).
\bibitem{R9}   A. G\7kalp and O. Y\5lmaz, Phys. Rev. {\bf D64}, 053017 (2001).
\bibitem{R10}   R. Escribano, Talk presented at the 9th International High-Energy Physics Conference in
Quantum Chromodynamics (QCD 2002), Montpellier, France, 2-9 July
2002, hep-ph/0209375 (2002).
\bibitem{R11}   Particle Data Group, D. E. Groom et al., Eur. Phys. J.{\bf C15}, 1 (2000).
\bibitem{R12}   V. E. Markushin, Eur. Phys. J. {\bf A8}, 389 (2000).
\bibitem{R13}   J. L. Lucio M., J. Pestieau, Phys. Rev. {\bf D42}, 3253 (1990); {\bf D43}, 2447 (1991).
\bibitem{R14}   M. N. Achasov et al., Phys. Lett.  {\bf B485}, 349 (2000).
\bibitem{R15}   E791 Collaboration,  E. M. Aitala et al., Phys. Rev. Lett. {\bf 86}, 770 (2001).
\bibitem{R16}   N. N. Achasov and V. V. Gubin, Phys. Rev. {\bf D63}, 094007 (2001).
\bibitem{R17}   S. M. Flatt\'{e}, Phys. Lett. {\bf B63}, 224 (1976).
\bibitem{R18}   A. Bramon, R. Escribano, J. L. Lucio M., M. Napsuciale, G. Pancheri, Phys. Lett. {\bf B494}, 221 (2000).
\bibitem{R19}   P. Singer, Phys. Rev. {\bf 128}, 2789 (1962); {\bf 130}, 2441 (1963); {\bf 161}, 1694 (1967).
\bibitem{R20}   A. Bramon, G. Colangelo and M. Greco, Phys. Lett. {\bf B287}, 263 (1992).

\end{thebibliography}
\end{document}